\documentclass[twocolumn,showpacs,preprintnumbers,amsmath,amssymb,aps,prl,reprint]{revtex4}
\usepackage{graphicx}
\begin{document}

\title{
Drive Dependence of the Skyrmion Hall Effect in Disordered Systems   
} 
\author{
C. Reichhardt and C. J. Olson Reichhardt 
} 
\affiliation{
Theoretical Division and Center for Nonlinear Studies,
Los Alamos National Laboratory, Los Alamos, New Mexico 87545, USA
} 

\date{\today}
\begin{abstract}
Using a particle-based simulation model, we show that quenched disorder 
creates a drive-dependent skyrmion Hall effect as measured by the change in the ratio 
$R=V_{\perp}/V_{||}$ of the skyrmion velocity perpendicular ($V_{\perp}$) and
parallel ($V_{||}$) to an external drive.
$R$ is zero at depinning and increases linearly with increasing drive, in agreement
with recent experimental observations.
At sufficiently high drives where
the skyrmions enter a free flow regime,
$R$ saturates to the disorder-free limit.
This behavior is 
robust for a wide range of disorder
strengths and intrinsic Hall angle values,
and occurs whenever plastic flow is present.
For systems with small
intrinsic Hall angles, we find that the
Hall angle increases linearly with external drive, as also observed in experiment. 
In the weak pinning regime where the skyrmion lattice depins elastically,
$R$ is nonlinear and
the net direction of the skyrmion lattice
motion can rotate as a function of external drive.      
\end{abstract}
\maketitle

Skyrmions in magnetic systems are particle-like objects 
predicted to occur in systems with chiral interactions \cite{1}.
The existence of a hexagonal skyrmion lattice in chiral magnets was
subsequently confirmed in neutron scattering experiments 
\cite{2} and   
in direct  imaging experiments \cite{3}.
Since then, skyrmion states have been found in an increasing number of 
compounds \cite{4,5,6,7,8},  including materials
where skyrmions are stable at room temperature \cite{9,10,11,12,12N}.
Skyrmions can be set into motion by applying an  
external current \cite{14,15}, and effective skyrmion velocity versus
driving force curves can be calculated from   
changes in the Hall resistance \cite{26,27}
or by direct imaging of the skyrmion motion \cite{9,12N}. 
Additionally, transport curves can be studied
numerically with continuum and particle based models \cite{28,29,30,M,31}.
Both experiments and simulations show that  
there is a finite depinning threshold for skyrmion motion similar to that found for the 
depinning of current-driven vortex lattices in type-II superconductors \cite{32,33,34}.   
Since skyrmions have particle like properties and can be moved with very low driving
currents,
they are promising candidates for spintronic applications \cite{35,36},
so an understanding of
skyrmion motion and depinning is of paramount importance. 
Additionally,
skyrmions represent an interesting dynamical system to study due   
to the strong non-dissipative
effect of the Magnus force they experience,
which is generally very weak or absent altogether in  
other systems where depinning and sliding phenomena occur. 

For particle-based representations of the motion of objects such as superconducting
vortices,
a damping term of strength
$\alpha_{d}$ aligns the particle velocity
in the direction of the net force acting on the particle,
while
a Magnus term of strength $\alpha_{m}$ rotates the   
velocity component in the direction perpendicular to the net force.
In most systems studied to date, the Magnus term is very weak compared to the
damping term, but
in skyrmion systems the ratio of the Magnus and damping terms can be as large
as 
$\alpha_{m}/\alpha_{d} \sim 10$  \cite{26,28,30,37}.
One consequence of the
dominance of the Magnus term is that under an external driving force, skyrmions
develop velocity components both
parallel ($V_{||}$) and perpendicular ($V_{\perp}$)
to the external drive, producing a skyrmion Hall angle of 
$\theta_{sk} = \tan^{-1}(R)$, where
$R=|V_{\perp}/V_{||}|$.
In a   
completely pin-free system, the intrinsic skyrmion Hall angle has a constant
value $\theta_{sk}^{\rm int} =  \tan^{-1}(\alpha_{m}/\alpha_{d})$;
however, in the  presence of pinning a moving skyrmion
exhibits a side jump phenomenon in the direction of the drive so  that the 
measured Hall angle is smaller than the clean value \cite{M,31,38}.
In studies of these side jumps 
using both continuum and particle based models for a skyrmion interacting with
a single pinning site \cite{M} and a periodic array of pinning sites \cite{38},
$R$ increases
with increasing external drive
until the skyrmions are moving fast enough that the pinning becomes ineffective and
the side jump effect is reduced.

In particle-based studies of skyrmions 
with an intrinsic Hall angle of $\theta^{\rm int}_{sk}=84^{\circ}$
moving through random pinning arrays,
$\theta_{sk}=40^{\circ}$ at small drives and increases with increasing drive
until saturating at $\theta_{sk}=\theta^{\rm int}_{sk}$ at
higher drives \cite{31}. 
In recent imaging experiments \cite{39} 
it was shown that $R=0$ and $\theta_{sk}=0$ at depinning and both
increase linearly with increasing drive; however,
the range of accessible driving forces 
was too low to permit observation of a saturation effect.  
These experiments were performed in a regime of relatively strong pinning,
where upper limits of $R \sim 0.4$ 
and $\theta_{sk} = 20^{\circ}$ are expected. 
A natural question is 
how universal the linear behavior of $R$ and $\theta_{sk}$ as a function of drive is,
and whether the results remain robust for larger intrinsic values of $\theta_{sk}$. 
It is also interesting to ask what happens in the weak pinning
limit where the skyrmions form a hexagonal lattice and
depin elastically.
In studies of overdamped systems such as superconducting vortices,
it is known that the strong and weak pinning
limits are separated by a transition from elastic to plastic depinning
and have very different transport curve characteristics \cite{32,34},
so a similar phenomenon could arise 
in the skyrmion Hall effect.   

{\it Simulation and System---} 
We consider a 2D simulation with periodic boundary 
conditions in the $x$ and $y$-directions
using a particle-based model of a modified Thiele equation 
recently developed for skyrmions interacting with
random \cite{30,31} and periodic
\cite{38,40} pinning substrates. 
The simulated region contains
$N$ skyrmions, and the time evolution 
of a single skyrmion $i$ is governed by the following equation: 
\begin{equation}  
\alpha_d {\bf v}_{i} + \alpha_m {{\bf \hat z}} \times {\bf v}_{i} = 
{\bf F}^{ss}_{i} + {\bf F}^{sp}_{i} + {\bf F}^{D}  .
\end{equation} 
Here, the skyrmion velocity is ${\bf v}_{i} = {d {\bf r}_{i}}/{dt}$,
$\alpha_d$ is the damping term, 
and $\alpha_m$ is the Magnus term. We impose the condition 
$\alpha_d^2 + \alpha_m^2 = 1$  
to maintain a constant magnitude of the skyrmion velocity  
for varied $\alpha_m/\alpha_d$.  
The repulsive skyrmion-skyrmion interaction force is given by
${\bf F}^{ss}_{i} = \sum^{N}_{j=1} \hat{\bf r}_{ij} K_{1}(r_{ij})$ 
where $r_{ij}=|{\bf r}_i - {\bf r}_j|$, 
$\hat{\bf r}_{ij}=({\bf r}_i - {\bf r}_j)/r_{ij}$,  
and $K_{1}$ is the modified Bessel function
which falls off exponentially for large $r_{ij}$.
The pinning force ${\bf F}^{sp}_{i}$
arises from non-overlapping randomly placed pinning sites modeled as
harmonic traps with an  amplitude
of $F_{p}$ and a radius of $R_{p} = 0.3$ as used in previous studies \cite{31}. 
The driving force 
${\bf F}^{D}=F_D {\bf \hat x}$ 
is from an applied current 
interacting with the 
emergent magnetic flux carried by the skyrmion \cite{26,37}. 
We increase
${\bf F}^{D}$ slowly to avoid transient effects.
In order to match the experiments, we take the driving force
to be in the positive $x$-direction so that
the Hall effect is in the negative $y$-direction. 
We measure the average skyrmion
velocity $V_{||}=\langle N^{-1}\sum_i^N {\bf v}_i \cdot {\bf \hat x}\rangle$
($V_{\perp}=\langle N^{-1}\sum_i^N {\bf v}_i \cdot {\bf \hat y}\rangle$)
in the direction parallel (perpendicular) to the applied drive,
and we characterize the Hall effect by measuring
$R  = |V_{\perp}/V_{||}|$ for varied $F^{D}$. 
The skyrmion Hall angle is
$\theta_{sk} = \tan^{-1}{R}$. 
We consider a system of size $L=36$ 
with a fixed skyrmion density
of $\rho_{sk}=0.16$ and pinning densities
ranging from $n_p=0.00625$ to  $n_p=0.2$. 

\begin{figure}
\includegraphics[width=\columnwidth]{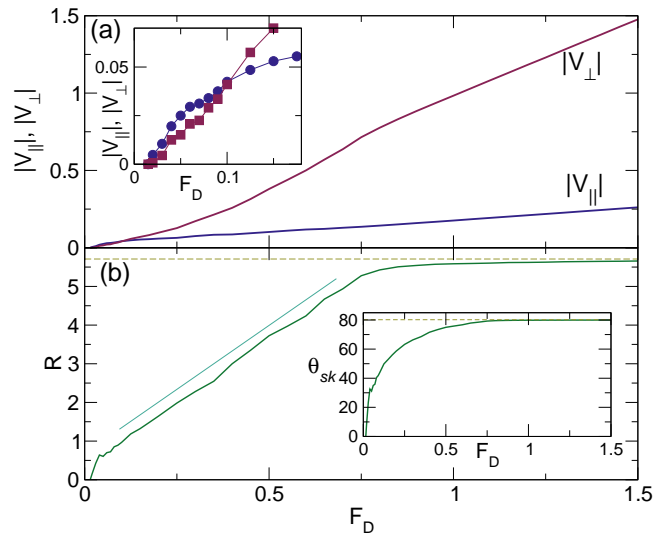}
\caption{
  (a) The skyrmion velocities in the directions parallel ($|V_{||}|$, blue)
  and perpendicular ($|V_{\perp}|$, red) to the driving force
  vs $F_{D}$
  in a system with $\alpha_{m}/\alpha_{d} = 5.71$, $F_{p} = 1.0$, and $n_{p} = 0.1$.
  The drive is applied in the $x$-direction.
  Inset: a blowup of the main panel in the region
  just above depinning where there is a crossing of the velocity-force curves. 
  (b) The corresponding $R = |V_{\perp}/V_{||}|$ vs $F_{D}$.
  The solid straight line is a linear fit 
  and the dashed line is the clean limit value of
  $R=5.708$.
  Inset: $\theta_{sk} = \tan^{-1}(R)$ vs $F_{D}$.    
The dashed line is the clean limit value of $\theta_{sk}=80.06^{\circ}$. 
}
\label{fig:1}
\end{figure}

{\it Results and Discussion---}
In Fig.~\ref{fig:1}(a,b) we  
plot $|V_{\perp}|$, $|V_{||}|$, and $R$  versus $F_{D}$ 
for  a system with 
$F_{p} = 1.0$, $n_{p} = 0.1$,
and $\alpha_{m}/\alpha_{d} = 5.708$.
In this regime, plastic depinning occurs, meaning that
at the depinning threshold some skyrmions
can be temporarily trapped at pinning
sites while other skyrmions move around them.  
The velocity-force curves are nonlinear, 
and $|V_{\perp}|$ increases more rapidly with increasing $F_D$ 
than $|V_{||}|$.
The inset of Fig.~\ref{fig:1}(a)
shows that 
$|V_{||}|>|V_{\perp}|$
for $F_D<0.1$, indicating 
that just above the depinning transition
the skyrmions are moving predominantly
in the direction of the driving force. 
In Fig.~\ref{fig:1}(b), $R$
increases linearly with increasing $F_D$ for  $0.04 < F_{D} <  0.74$, as
indicated by the linear fit,
while for $F_{D} > 0.74$
$R$ saturates to the intrinsic
value of $R=5.708$ marked with a dashed line.
The inset of Fig.~\ref{fig:1}(b) shows the
corresponding $\theta_{sk}$ vs $F_{D}$.
Initially $\theta_{sk}=0^\circ$, but $\theta_{sk}$ increases with increasing $F_D$ before
saturating at the
clean limit value of $\theta_{sk}=80.06^{\circ}$.
Although the linear increase in $R$ with $F_D$ is similar to the behavior observed
in the experiments of Ref.~\cite{39}, 
$\theta_{sk}$ does
not show the same linear behavior as in the experiments; 
however, we  
show later that when the intrinsic skyrmion Hall angle
is small, $\theta_{sk}$ varies linearly with drive.

\begin{figure}
\includegraphics[width=\columnwidth]{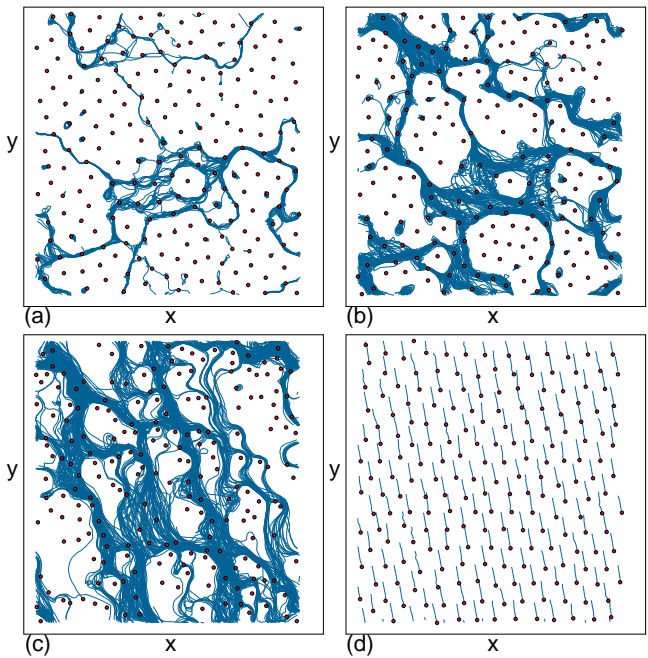}
\caption{ Skyrmion positions (dots) and trajectories (lines) obtained
  over a fixed time period
  from the system in Fig.~\ref{fig:1}(a).  The drive
is in the positive $x$-direction.  
(a) At $F_{D} = 0.02$, $R= 0.15$ and the motion is mostly
along the $x$ direction.
(b) At $F_{D} = 0.05$, $R = 0.6$ and the flow channels begin tilting into the
$-y$ direction.
(c) At $F_{D} = 0.2$, $R = 1.64$ and the channels tilt further
toward the $-y$ direction.
(d)  Trajectories obtained over a shorter time period
at $F_{D} = 1.05$ where $R  = 5.59$.  The skyrmions are dynamically 
ordered and move at an angle of $-79.8^{\circ}$ to the drive.
}
\label{fig:2}
\end{figure}

In Fig.~\ref{fig:2} we illustrate the skyrmion positions and trajectories obtained
during a fixed period of time at different drives
for the system in Fig.~\ref{fig:1}.
At $F_{D} = 0.02$ in 
Fig.~\ref{fig:2}(a),
$R = 0.15$ and the 
average drift is predominantly along the $x$-direction parallel to the drive, taking
the form of 
riverlike channels along which individual skyrmions intermittently
switch between pinned and moving states.
In Fig.~\ref{fig:2}(b),
for $F_{D} = 0.05$ we find $R=0.6$,
and observe wider channels that begin to tilt along the negative $y$-direction.
At $F_{D} = 0.2$ in
Fig.~\ref{fig:2}(c),
$R = 1.64$ and $\theta_{sk} = 58.6^{\circ}$.
The skyrmion trajectories are more strongly tilted along the $-y$ direction, and
there are still regions of temporarily pinned skyrmions
coexisting with moving skyrmions.
As the drive increases, 
individual skyrmions spend less time in the pinned state.
Figure~\ref{fig:2}(d) shows a snapshot of the trajectories over
a shorter time scale at 
$F_{D} = 1.05$
where $R = 5.59$. Here
the plastic motion is lost and the skyrmions form a moving crystal translating
at an angle of
$-79.8^{\circ}$ with respect to the external driving direction,
which is close to the clean value limit of $\theta_{sk}$.
In general, the deviations from linear
behavior that appear as $R$ reaches its saturation value in Fig.~\ref{fig:1}(b)
coincide with the loss of coexisting pinned and moving skyrmions, and are thus
correlated with the end of plastic flow.

\begin{figure}
\includegraphics[width=\columnwidth]{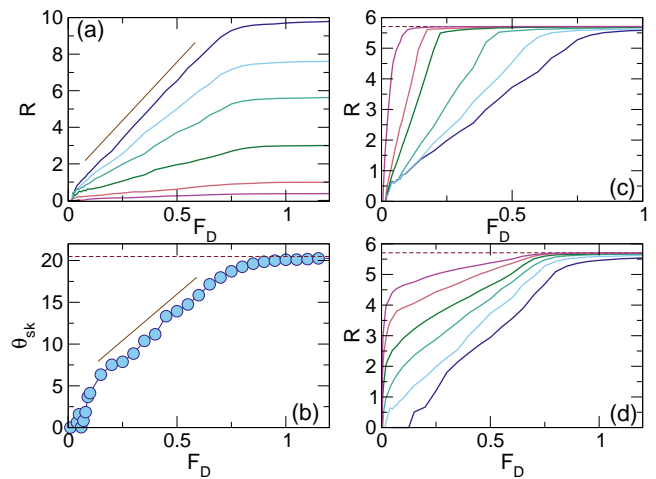}
\caption{ 
  (a) $R$ vs $F_{D}$ for samples with $F_{p} = 1.0$
  and $n_{p} = 0.1$ at
  $\alpha_{m}/\alpha_{d} = 9.962$,
  7.7367, 5.708, 3.042, 1.00, and $0.3737$, from left to right.  The line indicates 
  a linear fit.
  (b) $\theta_{sk} = \tan^{-1}(R)$ for $\alpha_{m}/\alpha_{d} = 0.3737$ from panel (a).
  The solid line is a linear fit 
and the dashed line indicates the clean limit value of $\theta_{sk} = 20.5^{\circ}$. 
(c) $R$ vs $F_{D}$ for $\alpha_{m}/\alpha_{d} = 5.708$
at $F_{p} = 0.06125$, 0.125, 0.25, 0.5, 0.75, and $1.0$, from left 
to right.
(d) $R$ vs $F_{D}$ for $F_{p} = 1.0$ at 
$\alpha_{m}/\alpha_{d} = 5.708$ for
$n_{p} = 0.00617$, 0.01234, 0.02469, 0.04938, 0.1, and $0.2$, from left to right.  
The clean limit value of $R$ is indicated by the dashed line.} 
\label{fig:3}
\end{figure}

In Fig.~\ref{fig:3}(a) we show $R$ versus $F_{D}$ for the system from
Fig.~\ref{fig:1} at varied 
$\alpha_{m}/\alpha_{d}$.
In all cases,
between the depinning
transition and the free flowing phase
there is a plastic flow phase
in which $R$ increases
linearly with $F_{D}$ with a slope that
increases with increasing $\alpha_{m}/\alpha_{d}$.
In contrast to the nonlinear dependence of $\theta_{sk}$ on $F_D$
at $\alpha_m/\alpha_d=5.71$ illustrated in the inset of Fig.~\ref{fig:1}(b),
Fig.~\ref{fig:3}(b) shows that for $\alpha_{m}/\alpha_{d} = 0.3737$,
$\theta_{sk}$
increases linearly with
$F_{D}$,
in agreement with the experiments of Ref.~\cite{39}.
Here,
$\theta_{sk}^{\rm int}= 20.5$,
close to the
value predicted in  the experiments of
Ref.~\cite{39}.
To understand the linear behavior,
consider the expansion of $\tan^{-1}(x) = x - x^{3}/3 + x^{5}/5 ...$
For small $\alpha_{m}/\alpha_d$,
as in the experiments,
$\tan^{-1}(R) \sim R$,
and since $R$ increases linearly with $F_D$, 
$\theta_{sk}$ also increases linearly with $F_D$.
In general, for $\alpha_{m}/\alpha_{d} < 1.0$ we find an extended region
over which
$\theta_{sk}$ grows linearly with $F_{D}$,
while for $\alpha_m/\alpha_{d} > 1.0$, the dependence of $\theta_{sk}$
on $F_D$ has nonlinear features similar to those shown in the inset of Fig.~\ref{fig:1}(b). 
In Fig.~\ref{fig:3}(c) we plot $R$ versus $F_{D}$ for a system 
with $\alpha_{m}/\alpha_{d} = 5.708$ for 
varied $F_{p}$.
In all cases $R$ increases linearly with $F_D$ before saturating; however,
for increasing $F_p$, the slope of $R$
decreases while the
saturation of $R$ shifts to higher values of $F_D$.
In general, the linear behavior
in $R$ is present whenever $F_p$ is strong enough to produce plastic flow.
In Fig.~\ref{fig:3}(d) we show 
$R$ versus $F_{D}$ at $\alpha_{m}/\alpha_{d} = 5.708$ for varied pinning densities
$n_{p}$.
In each case, there is a region in which
$R$ increases linearly with $F_{D}$,
with a slope that
increases with increasing $n_{p}$.
As $n_p$ becomes small, the nonlinear region just above depinning
where $R$ increases very rapidly with
becomes more prominent.

\begin{figure}
\includegraphics[width=\columnwidth]{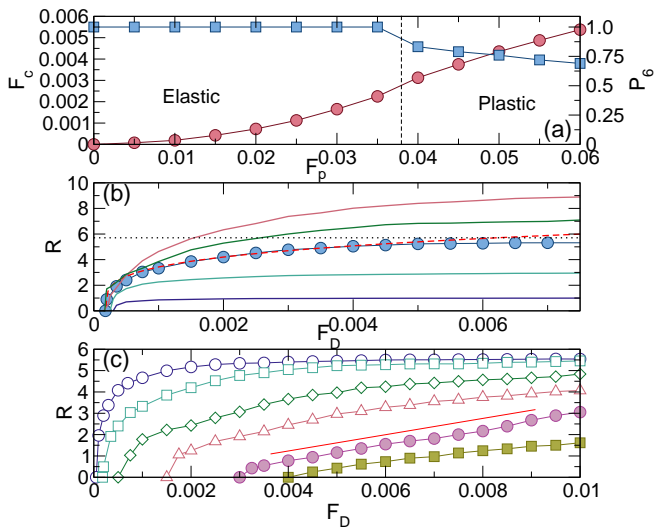}
\caption{
(a) Depinning force $F_{c}$ (circles)  and fraction $P_6$ of six-fold
coordinated particles (squares) vs $F_{p}$ for
a system with $\alpha_{m}/\alpha_{d} = 5.708$ and $n_p=0.1$,
showing a crossover from elastic depinning for $F_{p} < 0.04$
to plastic depinning for $F_{p} \geq 0.04$.
(b) $R$ vs $F_{D}$ for a system in the elastic depinning regime with
$F_{p} = 0.01$ and $n_{p} = 0.1$ at
$\alpha_{m}/\alpha_{d} =  9.962$, 5.708, 3.042, and $1.00$,
from top to bottom.  Circles indicate
the case $\alpha_{m}/\alpha_{d} = 5.708$,  for which
the dashed line is a fit to $R \propto (F_{D} - F_{c})^\beta$ with $\beta=0.26$ and
the dotted line indicates the pin-free value of $R = 5.708$. 
(c) $R$ vs $F_{D}$  for samples with $\alpha_{m}/\alpha_{d} = 5.708$ at
$F_{p} = 0.005$, 0.01, 0.02, 0.03, 0.04, and $0.05$, from left to right. 
The solid symbols correspond to values of $F_p$ for which plastic flow occurs, while
open symbols indicate elastic flow.
The line shows a linear dependence of $R$ on $F_D$ for
$F_{p} =0.04$.
}
\label{fig:4}
\end{figure}

For weak pinning, the skyrmions form a triangular lattice and exhibit
elastic depinning, in which each skyrmion maintains the same neighbors over time.
In Fig.~\ref{fig:4}(a) we plot the critical depinning force $F_{c}$
and the fraction $P_6$ of sixfold-coordinated skyrmions
versus $F_{p}$ 
for a system with $n_{p} = 0.1$ and $\alpha_{m}/\alpha_{d} = 5.708$. 
For $0 < F_{p} < 0.04$, the skyrmions depin elastically.
In this regime, $P_{6} = 1.0$ and $F_{c}$ increases as
$F_c \propto F_{p}^{2}$ as expected
for the collective depinning of elastic lattices \cite{33}.
For $F_{p} \geq 0.04$, $P_{6}$ drops
due to the appearance of
topological defects in the lattice,
and the system depins plastically,
with $F_{c} \propto F_{p}$ as expected for single particle
depinning or plastic flow.      

In Fig.~\ref{fig:4}(b) we plot $R$ versus $F_{D}$
in samples with $F_{p} = 0.01$ in the elastic depinning regime
for varied $\alpha_{m}/\alpha_{d}$.
We highlight the nonlinear behavior for the $\alpha_{m}/\alpha_{d} = 5.708$
case by
a fit of the form $R \propto (F_{D} - F_{c})^\beta$
with $\beta = 0.26$ and $F_{c} = 0.000184$.
The dotted line indicates the corresponding clean limit value of $R=5.708$. 
We find that
$R$ is always nonlinear
within the elastic flow regime,
but that there is no universal value of $\beta$,
which ranges from
$\beta=0.15$ to $\beta=0.5$
with varying $\alpha_{m}/\alpha_{d}$.
The change in the Hall angle with drive is most pronounced
just above the depinning threshold, as indicated by 
the rapid change in $R$ at small $F_D$.
This
results from the elastic
stiffness of the skyrmion lattice which prevents individual skyrmions from occupying
the most favorable substrate locations.
In contrast, $R$ changes more slowly at small $F_D$ in the
plastic flow regime, where
the softer skyrmion lattice can adapt to the
disordered pinning sites.
In Fig.~\ref{fig:4}(c) we plot $R$ versus $F_{D}$ at
$\alpha_{m}/\alpha_{d} = 5.708$ for varied
$F_{p}$,
showing a reduction in $R$ with increasing $F_{p}$.
A fit of the $F_{p} = 0.04$ curve in the plastic depinning regime
shows a linear increase of $R$ with $F_{D}$,
while for $F_{p} < 0.04$ in the elastic regime, the dependence of $R$ on $F_D$ is
nonlinear.   
Just above depinning
in the elastic regime,
the skyrmion flow direction
rotates
with increasing drive.  

{\it Summary---} 
We have investigated the skyrmion Hall effect by measuring the ratio $R$ of the skyrmion
velocity perpendicular and parallel to an
applied driving force.  
In the disorder-free limit,
$R$ and the skyrmion Hall angle
take constant values 
independent of the applied drive; however, in the presence of pinning 
these quantities become drive-dependent, and in the  strong pinning regime
$R$
increases linearly
from zero
with increasing drive,
in agreement with recent experiments.
For large intrinsic Hall angles,
the current-dependent Hall angle increases nonlinearly with
increasing drive;
however, for small
intrinsic Hall angles
such as in recent experiments,
both the current-dependent Hall angle and
$R$
increase linearly with
drive as found experimentally.
The linear dependence of
$R$ on drive is robust for a wide range of intrinsic
Hall angle values, pinning strengths, and pinning densities,
and appears whenever the system exhibits plastic flow.
For     
weaker pinning forces where the skyrmions depin elastically,
$R$ has a nonlinear drive dependence and increases very rapidly
just above depinning.
We observe a crossover from nonlinear to linear drive dependence of
$R$
as a function of the pinning force, which coincides with the transition from
elastic to plastic depinning.

\begin{acknowledgments}
We gratefully acknowledge the support of the U.S. Department of
Energy through the LANL/LDRD program for this work.
This work was carried out under the auspices of the 
NNSA of the 
U.S. DoE
at 
LANL
under Contract No.
DE-AC52-06NA25396.
\end{acknowledgments}

\end{document}